\documentclass[12pt,a4]{article}

\usepackage{a4wide}
\usepackage[T1]{fontenc}
\usepackage{amsmath}
\usepackage{amssymb}
\usepackage{epsfig}
\begin{document}

\begin{center}
{\Large Nonlinear rheology of Laponite suspensions \\ under an external drive}\\
\vspace{0.5cm}
{\large B\'ereng\`ere Abou $^{ a) }$, Daniel Bonn,  and J. Meunier}\\

Laboratoire de Physique Statistique, UMR CNRS 8550, Ecole Normale Supérieure, \\
24, rue Lhomond, 75005 Paris, FRANCE\\
\end{center}

\noindent $^{a) }$ present address : Laboratoire de Biorhéologie et Hydrodynamique Physico-chimique,
UMR CNRS 7057, Université Paris VII, 4 Place Jussieu, 75005 Paris, FRANCE\\
abou@ccr.jussieu.fr

\begin{abstract}

We investigate the nonlinear rheological behavior of colloidal suspensions of Lapo\-nite, 
a synthetic clay, driven 
by a steady and homogeneous shear strain. We show that the external drive leads to a drastic slowing down of the aging dynamics or even, in some cases, in the rejuvenation of the system. Under shear, after a surprisingly long time, the spontaneous aging process observed at rest is suppressed. The system then reaches a non-equilibrium stationary state, characterised by a complex viscosity depending on the applied shear rate. In addition, the glass exhibits a non-Newtonian shear-thinning behavior. These rheological behaviors confirm recent numerical and theoretical predictions. 

\end{abstract}

\section{Introduction}

Complex fluids, such as foams, emulsions, pastes and slurries, are known to exhibit a rich phenomenology of 
nonlinear rheological behaviors (Larson, 1999). In particular, similar low frequency shear rheology is observed for a wide range of these soft materials (see Fielding {\it et al.}, 2000 for a review). For many of these materials, the stress response $\sigma $ to a shear strain of constant rate $\dot{\gamma}$ is described by $\sigma = A + B \dot{\gamma}^n$ that leads to a non-newtonian viscosity (Barnes {\it et al.}, 1989). This common rheological behavior was recently attributed to two properties of these materials, {\it structural disorder} and {\it metastability}, resulting in a ``glassy'' dynamics, where thermal 
motion alone is not sufficient to achieve 
complete structural relaxation (Sollich {\it et al.}, 1997; Sollich, 1998).  

Due to long-range interactions, it turns out to be a complex task to predict the nonlinear rheological behavior, including a non-Newtonian viscosity, on the basis of the structure and interactions in these systems. Numerical simulations have allowed to predict the relationship between the material's microstructure and its macroscopic properties (Brady {\it et al.}, 2000; Melrose {\it et al.}, 1995; 
Melrose {\it et al.}, 1999). 
Recently, a different approach to the rheology of these ``soft glassy materials'' has been developed, starting from a model of a glassy system instead of taking all the interactions -- including colloidal and hydrodynamic interactions -- into account. Because 
of both their practical and fundamental interest, the nonlinear rheology of such soft glassy systems subjected to an external drive has recently attracted particular attention. This was done either by extending approaches for the description of glassy dynamics (Sollich {\it et al.}, 1997; Cugliandolo {\it et al.}, 1997; Sollich, 1998;  Berthier {\it et al.}, 2000; Barrat and Berthier, 2000; Fielding {\it et al.}, 2000) or based on phenomenological approaches (H\'ebraud {\it et al.}, 1998; D\'erec {\it et al.}, 2001, Coussot {\it et al.}, 2002). The most 
interesting situation occurs when the relaxation time scale of the system becomes of the same order of magnitude as the time 
scale introduced by the shearing process. This usually happens in glassy systems because the typical relaxation 
time grows together with the waiting time $t_w$, that is the time elapsed after a quench from the ``liquid'' into the ``glassy'' state. 

At rest, the properties of glassy systems thus depend on the waiting time: the system is said to {\it age}. Recent theoretical and numerical works, initially developped 
for supercooled liquids, have allowed for a first detailed description of the aging process (Bouchaud {\it et al.}, 1996; Kob and Barrat, 1997). Glassy systems exhibit two modes 
of relaxation : a fast mode corresponding to the fast motion of particles in the 'cages' 
constituted by their neighbors and a slow mode 
of relaxation resulting from the structural rearrangement of these 'cages'. The strongest 
experimental evidence for the applicability 
of such predictions (including classical mode-coupling approach) comes from colloidal 
glasses (Nelson and Allen, 1995; Pusey and Van Megen, 1987; Bonn {\it et al.}, 1999; 
Abou {\it et al.}, 2001). Because of the experimental advantages they present compared to 
structural glasses, where there are two coupled control parameters, temperature and density, these 
colloidal glasses have been studied extensively. 

In this paper, we investigate experimentally the response of a colloidal glass of Laponite to a steady shear strain of constant rate. The experimental materials and methods are described in Section II. Section III presents the rheological experiments. We discuss our experimental results and conclude in Section IV.

\section{Materials and methods}

\begin{figure}
\center{\epsfig{file=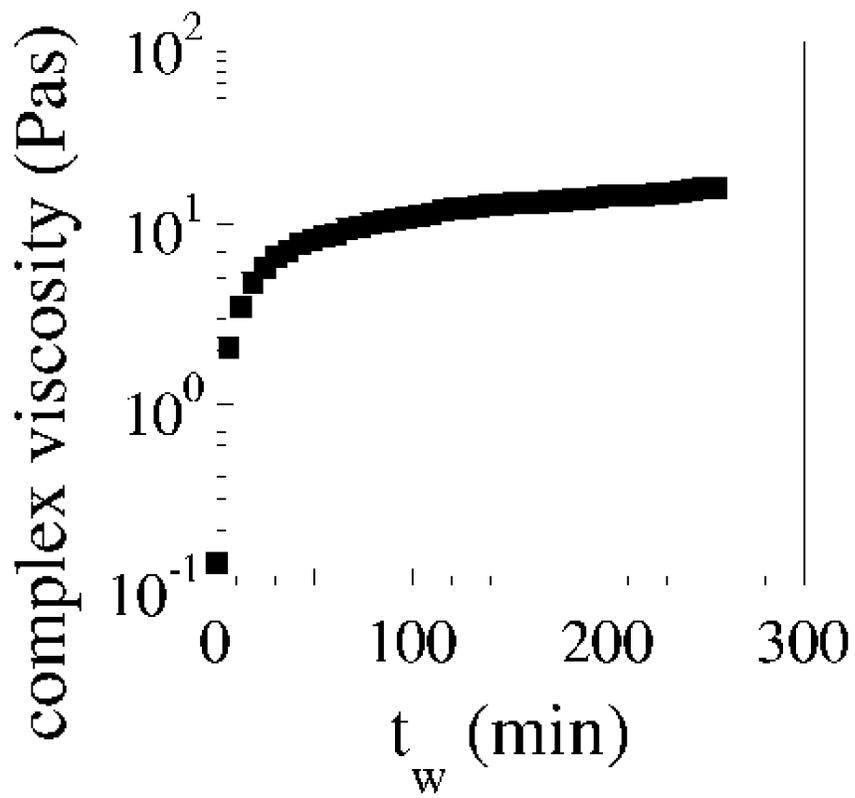, width=12cm}}
\caption{Complex viscosity as a function of the aging time $t_w$ for a Laponite suspension at $1.5 \%$ wt, $I = 7 . 10^{-3}$ M.}
\label{viscosite}
\end{figure}

The experiments were performed on Laponite RD, a synthetic clay manufactured by Laporte Industry. The chemical composition of the clay is as follows : SiO$_2$, $65.82 \%$; MgO, $30.15 \%$; Na$_2$O, $3.20 \%$; LiO$_2$, $0.83 \%$ and corresponds to the chemical formula Si$_{8.00}$ (Mg$_{5.45}$Li$_{0.40}$)H$_4$O$_{24}$Na$_{0.75}$. The particles of Laponite are colloidal disks of 25 nm diameter and 1 nm thickness, with a negative surface charge on both faces (Kroon {\it et al.}, 1998). The clay powder was mixed in ultra-pure water and NaCl at different clay concentrations varying between $2 \%$ wt to $5 \%$ wt. The ionic strength of the suspension, denoted $I$, was subsequently adjusted by dissolution of NaCl in water, in the range $ 10^{-4 } < I < 8. 10^{-3} $ M. The pH value of the suspensions was fixed to pH = 10 by addition of NaOH, providing chemically stable particles (Thompson and Butterworth, 1992; Mourchid and Levitz, 1998). The suspension was stirred vigorously during 15 minutes and then filtered through a Millipore Millex - AA 0.8 $\mu$m filter unit. This preparation procedure allows us to obtain a reproducible initial liquid state. The aging time $t_w = 0$ of the suspension is defined as the moment it passes through the filter. Within a time varying from a few minutes to a few hours for the different concentrations and ionic strengths considered here, a three order of magnitude increase in the suspension viscosity was observed. Being in a liquid state right after preparation, the suspension becomes more and more viscous and does not flow anymore when tumbling the recipient. Since the physical properties of the suspension depend on the time elapsed since preparation $t_w$, the sample is said to age. The aging dynamics can be characterised by oscillatory shear experiments in order not to disturb the system. These measurements are performed using small oscillations at a frequency of 1 Hz at an imposed stress of 0.1 Pa. They yield G' and G'', the elastic and loss moduli respectively. The complex viscosity modulus $\eta ^*$ can be calculated from these quantities from $\eta^*= (G'^2 +G''^2)^{1/2} / \omega$. It can be observed that the viscosity changes very rapidly, by 2 orders of magnitude over about 1 hour, as shown in Figure \ref{viscosite}.  
\begin{figure}
\center{\epsfig{file=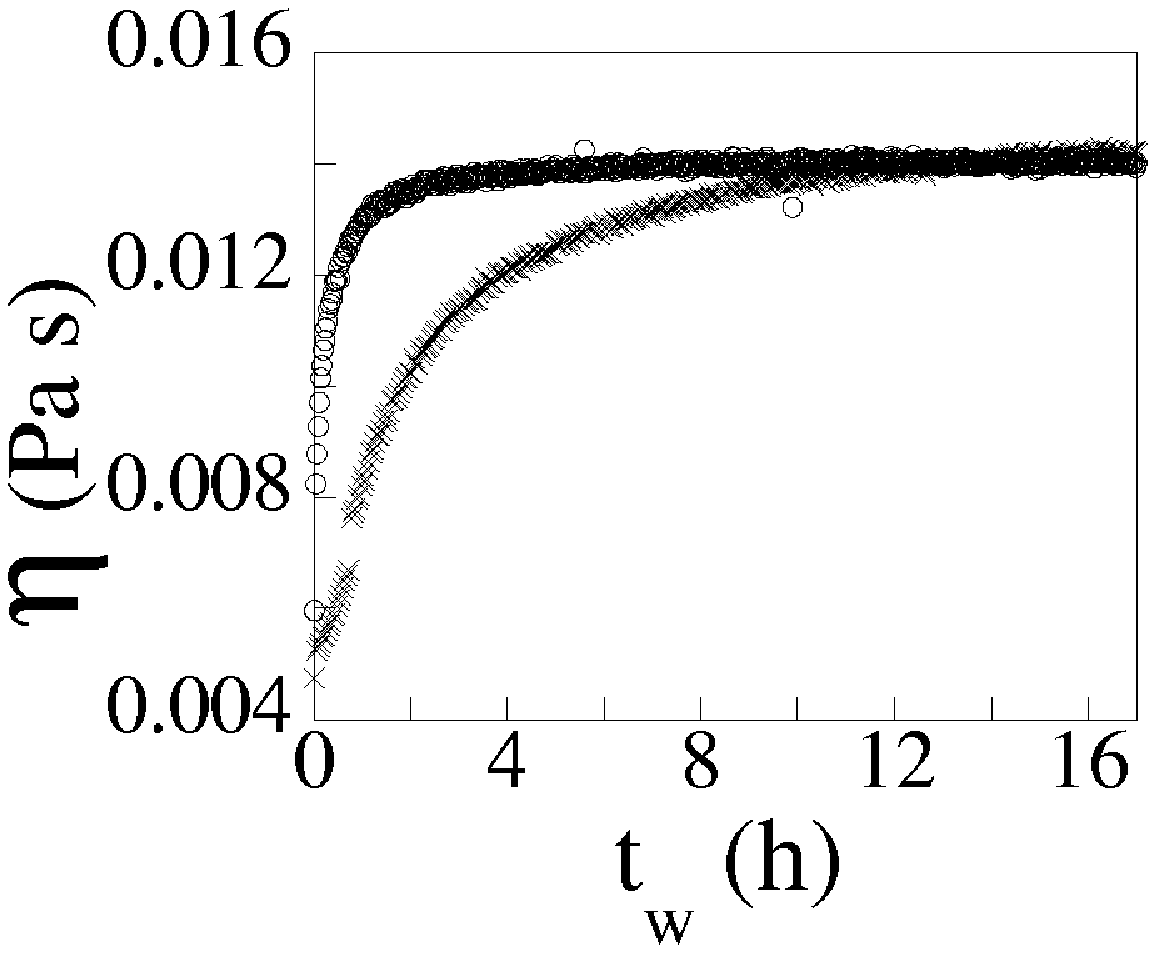, width=12cm}}
\caption{Effect of the addition of NaCl on the driven aging dynamics : viscosity as a function of the aging time $t_w$ for Laponite suspensions at $1.5 \%$ wt, $I = 5 . 10^{-3} $ M  (cross) and $I = 7 . 10^{-3}$ M (circles), by applying a constant shear rate $\dot{\gamma}= 500$~s$^{-1} $.}
\label{sel}
\end{figure}
Thus, suspensions of Laponite become strongly visco-elastic in time, even at low volume fraction $\Phi$ in particles (of the order of $\Phi \simeq 0.01$). The formation of a gel, evidenced by the existence of a fractal network, has been invoked in explaining the visco-elasticity (Pignon {\it al.}, 1997; Mourchid and Levitz, 1998). Recently, the structure and viscosity of Laponite suspensions at ionic strength $ I = 10^{-4}$  M were studied by using static light scattering (Bonn {\it et al.}, 1999). Contrary to previous observations, no evidence for a fractal-like organisation of the particles could be found, provided the suspensions were filtered. Therefore, it has been proposed that Laponite suspensions form so-called repulsive colloidal glasses, stabilised by electrostatic repulsions. In such systems, the ergodicity is lost due to blocking of particle movements by the dense surrounding cages formed by their nearest 
neighbours. The colloidal glass is obtained for very low volume fraction $\Phi \simeq 0.01$ compared to those for usual spherical colloids, for which glasses are obtained above $\Phi \simeq 0.5 $ (Pusey and Van Megen, 1987). To account for this difference, one has to consider the effective volume fraction of a particle, which is done by adding to the particle radius the Debye length. To give an order of magnitude, it is estimated to 30 nm for a ionic strength $I = 10^{-4} $ M. Recent experiments have shown that the location of the glass transition line in the (volume fraction / electrolyte concentration) phase diagram is consistent with this assumption (Levitz {\it et al.}, 2000). 

Addition of salt to a suspension of Laponite disks reduces electrostatic repulsion and can even lead to an attraction. Recently, such short-ranged attractive colloidal systems have received renewed attention 
in terms of their dynamic properties (Nicola\"{\i} and Cocard, 2001). When the strength of short-ranged attraction becomes 
significantly greater than the thermal energy, the system can form a colloidal gel, in which the particle motions will 
be completely jammed even at very low bulk volume fractions. The system 
behaves nonergodic prior to complete blocking of particle dynamics. Therefore, by varying the strength and range 
of the attractive 
interactions, a reentrant transition of the liquid glass line and a glass-glass transition can be realized (Dawson {\it et al.}, 2001). 
 
In this paper, we present mainly experimental results obtained with suspensions of Laponite at $1.5 \%$ wt, $I = 7 . 10^{-3}$ M. Very similar results were obtained with suspensions of Laponite at $I = 10^{-4}$ M. For the ionic strength considered here, static and dynamic light scattering experiments show that the suspensions form colloidal glasses (Bonn {et al.}, 1999).

\section{Nonlinear rheology}

Shear experiments were performed on a controlled stress rheometer that can also operate in controlled strain mode (Rheologica StressTech), using a Couette geometry or a vane geometry with 1 mm gap in the range $ 50 < \dot{\gamma}  <  500 $~s$^{-1}$, and a Couette geometry with 1 / 8 mm gap in the range $ 500 < \dot{\gamma} <  2000$  ~s$^{-1}$. The tests were carried out at a temperature of $20.0  \pm 0.5^ {\mbox {{\tiny{o}}}} $C. Evaporation of water or CO$_2$ contamination of the sample was completely avoided by covering the sample with vaseline oil and a plate made of Plexiglas.
 
At $t_w = 0$, the sample was exposed to a continuous shear strain with a constant rate $\dot{\gamma}$ between $50 < \dot{\gamma} <  2000$~ s$^{-1}$. The viscosity of the suspension increased slowly with time and reached a stationary value after some time, as shown in Figure \ref{sel}. A variation of the viscosity of less than $1 \%$ per hour was choosen as a criterion for the stationary state to be reached, as this is of the order of magnitude of our measurements precision. This stationary state was obtained after very long times. To give an order of magnitude, for a $4 \%$ wt suspension at ionic strengh $I = 10^{-4}$ M, it was reached in about 50 hours. By increasing the particle concentration from $1.5 \%$ to $5 \%$, we observed a slight decrease in the time to reach the stationary state. Upon increasing the ionic strength, we were able to reach the stationary state of the system faster, as shown in Figure \ref{sel} for a suspension at $1.5 \%$ wt, with two different ionic strengths. 

\begin{figure}
\center{\epsfig{file=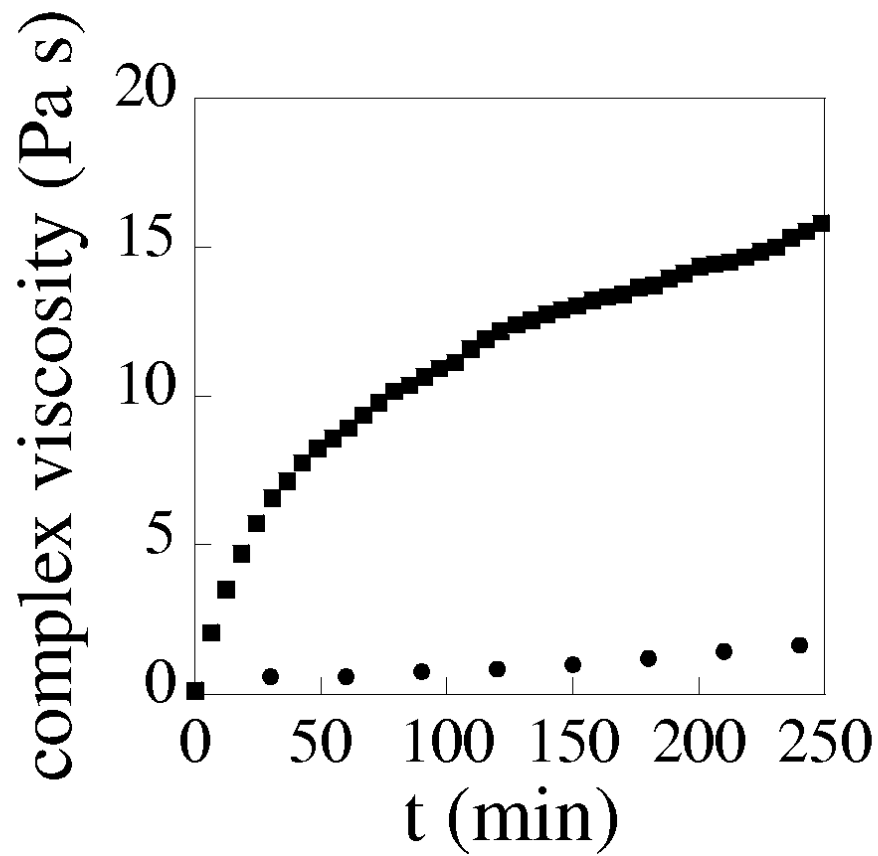, width=12cm }}
\caption{Comparison between the aging dynamics under shear (circle) and the spontaneous one (square) for a Laponite suspension at $1.5 \%$ wt, $I = 7 . 10^{-3} $ M: the measured complex viscosity modulus varies slowly when the system is submitted to an external drive ($\dot{\gamma}= 500$~s$^{-1} $) compared to spontaneous aging. }
\label{spont}
\end{figure}

In order to compare the driven aging dynamics to the spontaneous one, we measured the complex viscosity modulus of the suspension under shear. The experimental procedure was the following : every 30 minutes, we stopped shearing for a very short time (typically 30 s) and performed oscillatory shear experiments, at a frequency of 1 Hz and at an imposed stress of 0.15 Pa, leading to the elastic and loss moduli. This procedure allowed us no to perturb the shearing process. The complex viscosity moduli both for the system that evolves spontaneously and for the system under shear are shown in Figure \ref{spont}. As the complex viscosity modulus can be taken as a measure of the age of the system, it follows immediately that the aging dynamics under flow is very slow compared to the ``spontaneous'' aging dynamics : the complex viscosity modulus increase is slower when the system is submitted to an external drive. 
If the drive was suppressed once the stationary state was attained, the ``spontaneous'' aging process again took place as was observed by performing oscillatory shear experiments : the complex viscosity modulus $\eta^*$ changed very rapidly, by 2 orders of magnitude over about 1 hour, in a similar way to what was observed for the spontaneous aging process. 

Starting from different aging times $t_w$, a continuous shear strain with the same constant rate was applied to the system : the same non-equilibrium stationary state is attained, characterized by very similar viscosities. Figure 4 shows the rejuvenation and driven aging process under a shear rate $ \dot{\gamma} = 500 $~s$^{-1}$, starting from different aging times. The steady-state viscosity thus appears to only depend on the applied shear rate. As the complex viscosity can be taken as the measure of the age of the system, the corresponding age of the stationary state appears to be power-dependent. When applying the external drive to a suspension aged for a long enough time $t_w$, the viscosity decreased in time until the system reached its steady state. During this shearing experiment, we measured the complex viscosity modulus using the same procedure as explained above. The complex viscosity modulus and the dynamic viscosity are shown to evolve in the same direction, they both decrease : so-called rejuvenation of the suspension was observed. It is worth noting that, even by shearing strongly the suspension during hours, in the range $50 < \dot{\gamma} <  2000$~ s$^{-1}$, the initial ``liquid'' state, obtained at $t_w=0$ as described in Section II, was never reached again, the final state being only determined by the rate of the applied shear strain. 
\begin{figure}
\center{\epsfig{file=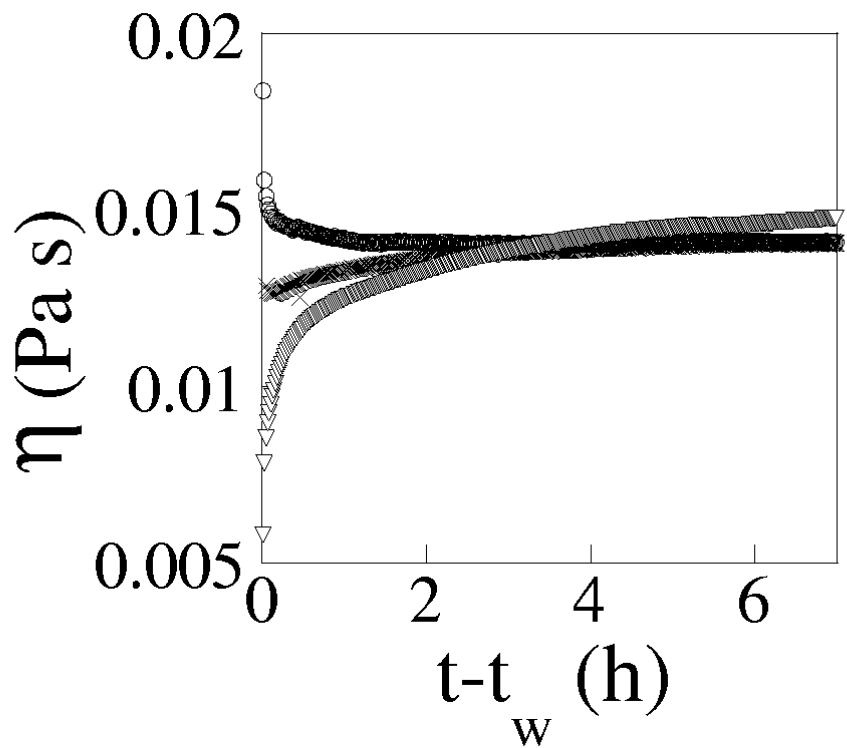, width=12cm}}
\caption{Aging dynamics under a shear strain with a constant rate 
$ \dot{\gamma} = 500 $~s$^{-1}$ in a Laponite suspension ($1.5 \%$ wt, $I = 7.10^{-3}$ M). Starting from different aging times $t_w = 0 $ (triangle), $ t_w = 20 $ min (cross), $ t_w = 40 $ min (circle), the glass exhibits so-called rejuvenation (circle) and driven aging dynamics (cross and triangle). The stationary state reached only depends on the rate of the applied shear strain. }
\label{stat}
\end{figure}

\begin{figure}
\center{\epsfig{file=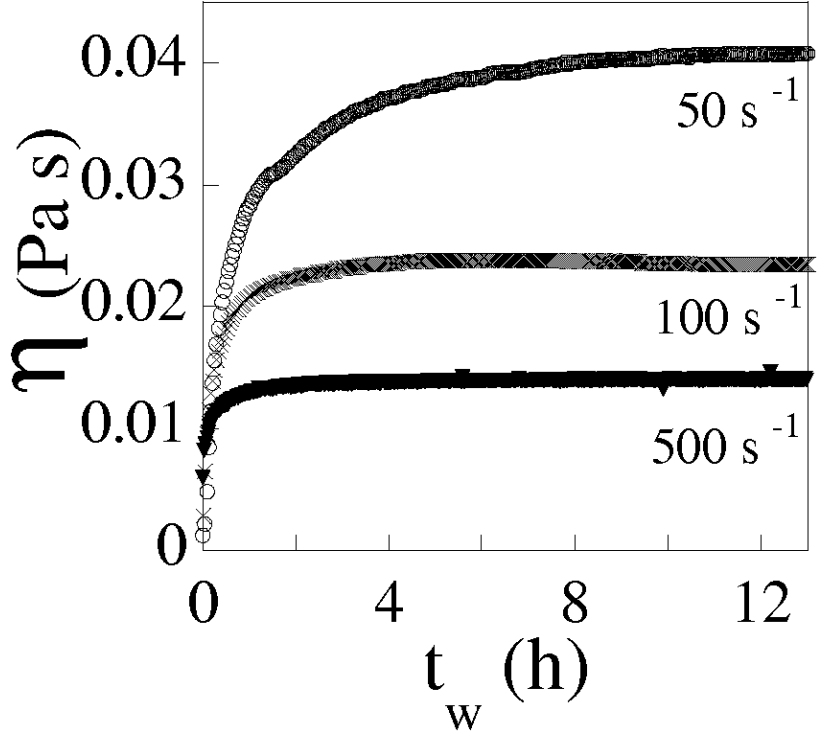, width=12cm}}
\caption{Viscosity as a function of the aging time $t_w$ for a Laponite suspension at $1.5 \% $ wt, $I = 7. 10^{-3}$ M, 
for various applied shear rates $50$~s$^{-1}$, $100 $~s$^{-1}$ and $ 500$~s$^{-1} $. The larger the applied shear rate, the smaller the value of the stationary viscosity and the faster to reach it.}
\label{diffgam}
\end{figure}

\begin{figure}
\center{\epsfig{file=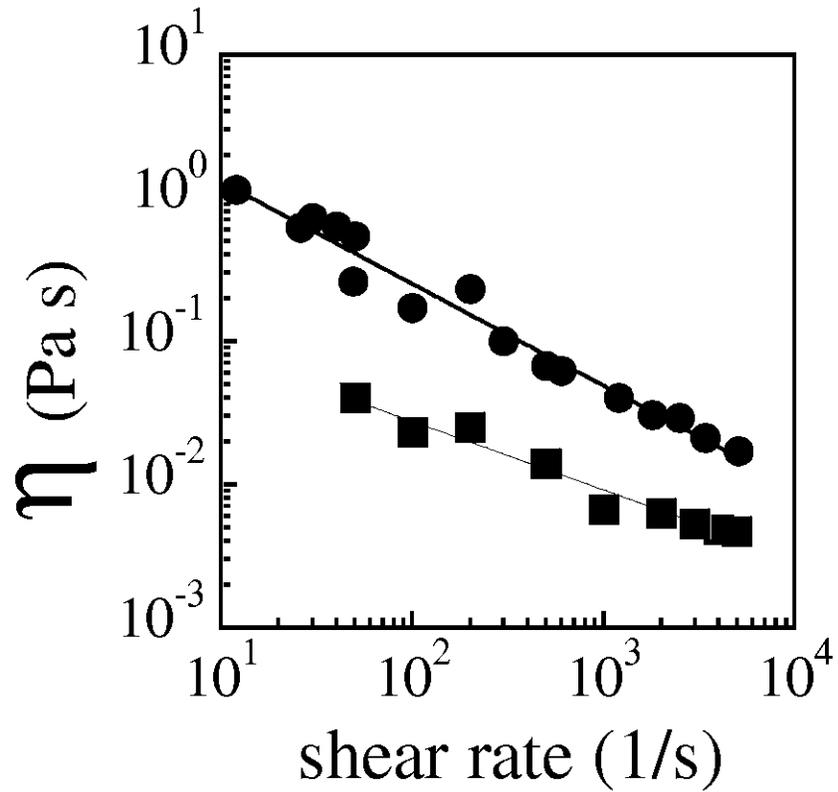, width=12cm}}
\caption{Stationary viscosity as a function of the applied shear rate for a suspension at $1.5 \%$ wt, $I = 7. 10^{-3}$ M (square) and a suspension at $3.7 \%$ wt, $I = 10^{-4}$ M (circle). We observe a shear-thinning behaviour characterized by a a power law $\eta \propto \dot{\gamma}^{- \alpha}$, with $\alpha = 0.60 \pm 0.10$.}
\label{rheof}
\end{figure}

The stationary viscosities obtained for different values of the applied shear rate from $t_w = 0$ are shown in Figure \ref{diffgam}. The larger the applied shear rate, the smaller the value of the stationary viscosity and the faster we reach it. If we consider the typical time necessary to obtain the steady state as the time when the viscosity varies less than $1\%$, we find that for the constant respective shear rates $50$~s$^{-1}$, $100 $~s$^{-1}$ and $ 500$~s$^{-1} $, times are respectively 10 h, 5 h and 1 h. As a result, it roughly scales with $1 / \dot{\gamma}$. By reporting the values of the stationary viscosity as a function of the applied shear rate, a shear-thinning behaviour was observed as shown in Figure \ref{rheof}. It can be accurately characterized by a power law $ \eta \propto \dot{\gamma}^{- \alpha} $, with $ \alpha = 0.60 \pm 0.10 $ for the different suspensions of Laponite considered here. \\

\section{Discussion}

From a theoretical point of view, a number of models have been introduced recently to account for 
the interplay between the drive and the relaxation of the system. Sollich and co-wokers (1997) extended Bouchaud 's trap model (Bouchaud, 1992; Monthus and Bouchaud, 1996) to driven systems, leading to the 'Soft Glassy Rheology' (SGR) model (Sollich, 1998; Fielding {\it et al}, 2000). Other models that capture many of the observed rheological properties have been introduced (H\'ebraud {\it et al.}, 1998; D\'erec {\it et al.}, 2001, Coussot {\it et al.}, 2002). Cugliandolo {\it et al.} (1997) suggested an approach to non-equilibrium systems, in which the non-equilibrium state was generated by 'stirring' the system. Further theoretical studies on driven mean-field disordered systems and numerical studies on Lennard-Jones glasses allowed for a detailed description of the non-equilibrium behaviour, including rheological properties (Berthier {\it et al.}, 2000; Barrat and Berthier, 2000). All these models were shown to account well for a number of rheological behaviors in soft glassy materials. In particular, as a result of these theoretical predictions, it is expected that if we inject power into aging systems, there is the possibility of stabilizing the age of the system in a power dependent level : the younger the larger the power input (Bouchaud {\it et al.}, 1996; Horner, 1996; Sollich, 1998; Kurchan, 1999; Berthier {\it et al.}, 2000; D\'erec {\it et al.}, 2001). The power injection into the system derives from the applied shear rate. For a vanishingly small drive, however, the slow mode relaxation time, assumed to be proportional to the viscosity, again diverges (Berthier {\it et al.}, 2000).

These theoretical predictions are confirmed by the experimental results described above. The driving force acting on the system results in the suppression of the aging process. The larger the shear rate, the smaller the stationary viscosity that means the younger the glassy system, as the complex viscosity is known to increase with time $t_w$ and can be therefore taken as a measure of its age. Once the drive is suppressed, the spontaneous aging process again takes place. From our experimental results, it appears that the external drive leads to a drastic slowing down of the spontaneous aging dynamics or, in some cases, in the rejuvenation of the system. The occurence of rejuvenation or ``driven'' aging depends on both the external drive power and the age of the system when applying the external drive. This is completely consistent with phenomenological models in which the final stationary state results from the competition between the spontaneous aging dynamics -- that strenghtens the interactions 
between particles and leads to less and less accessible configurations in phase space as time evolves -- and the mechanically induced rejuvenation of the configurations -- process going against the spontaneous aging and leading to its slowing down. Coussot and co-workers (2002) have shown that, under imposed stress a bifurcation in rheological behavior occurs: for small stresses, the viscosity increases in time: the material eventually stops flowing. For slightly larger stresses, the viscosity decreases continuously in time: the flow accelerates. Thus, the viscosity jumps discontinuously to infinity at the critical stress, implying also there is a range of shear rates for which the  flow is unstable. Here, we work at sufficiently high  imposed shear rates so that the flows are stable; no shear banding is observed. 

In the experiments, the typical time to reach the stationary state is surprisingly long. Our experiments show that this typical time scales with $ 1 / \dot{\gamma} $, result which confirms theoretical expectations and numerical simulations (Berthier {\it et al.}, 2001). However, the prefactor of $ 1 / \dot{\gamma} $ is of the order of $10^6$ in a typical experiment and suggests that the applied shear disrupts very large scale structures. We observed that this typical time is of the same order of magnitude than the time necessary for the spontaneous aging dynamics to slow down dramatically. This fact strengthens the idea of a competition, between spontaneous aging and the external drive. Furthermore, these steady-shear experiments disprove a common idea on Laponite suspensions : when taking an ``old'' suspension, it is not sufficient to stir the system during a few minutes in order to obtain the initial state at $t_w = 0$, for the shear strain rates considered here.

The shear rejuvenation was studied by using Diffusive Wave Spectroscopy (DWS) in a suspension of Laponite (Bonn {\it et al.}, 2002). The shear rejuvenation was shown to have a very large effect on the microscopic dynamics. Upon increasing the shear rate, the steady-state relaxation time decreases. In the experiments reported in this paper, the complex viscosity modulus associated to the stationary state was also shown to decrease with the applied shear rate. This is consistent with the DWS experiments, where the measurements allowed us to relate the microscopic diffusion dynamics to the macroscopic viscosity of the system.

In the SGR model, the rheological response of the system depends on the distance from the glass transition temperature $x$. If we assume that we are in the immediate vicinity of the glass transition, $x \simeq 1$, the shear-thinning exponent found experimentally is in reasonable agreement with the theory.  
Berthier {\it et al.} (2000) studied, within the framework of the mean-field (or mode-coupling) approximation, the influence of an external drive on a system undergoing a glass transition. Above the glass transition, for high enough shear rates, the slow mode relaxation time $t_{\alpha}$ of the system scales as $t_{\alpha} \propto \dot{\gamma}^{-2/3}$. As the dynamic viscosity $\eta$ is assumed to be proportional to the slow mode relaxation time, the system has therefore a shear-thinning behavior characterized by the power law $\eta \propto \dot{\gamma}^{-2/3}$. Very similar results are obtained from molecular dynamics simulations, where the viscosity is directly extracted from the applied shear rate (Barrat and Berthier, 2000; Yamamoto and Onuki, 1998), with $\eta \propto \dot{\gamma}^{-0.9}$ and $\eta \propto \dot{\gamma}^{-0.8}$ respectively. The shear-thinning behavior observed experimentally (an exponent between $-0.5$ and $-0.7$) is in fair agreement with the above theoretical and numerical predictions. 

In conclusion, we have studied experimentally the nonlinear rheological behavior of a glassy system under an external drive. Qualitatively, our experimental results confirm all the findings predicted by the glassy models. They show the competition between the external drive, leading to a mechanically induced rejuvenation of the configurations, and the aging dynamics, leading to a slowing down. Depending on the competitive antagonists, driven aging or rejuvenation may be observed. In addition, the shear-thinning viscosity observed in our experiments agrees quantitatively with at least one simulation result. Since the value of the exponent depends on the exact model that is chosen, the differences with the model calculations may not be very important. The key result is that non-equilibrium models for glassy systems are able to predict a non-Newtonian behavior without taking into account the specific interactions between particles. 

\newpage

\section{References}

Abou, B., Bonn, D., and Meunier, J., Aging dynamics in a colloidal glass of laponite, Phys. Rev. E. {\bf 64}, 021510 1-6 (2001).\\
Barnes, H. A., Hutton, J. F., and Walters, K., {\it An Introduction to Rheology} 
(Elsevier, Amsterdam, 1989).\\
Barrat, J.-L, and, Berthier, L., Fluctuation-dissipation relation in a sheared fluid, Phys.Rev. E {\bf 63}, 012503 1-4 (2000).\\
Berthier, L., Barrat, J.-L., and Kurchan, J., A two-time-scale, two-temperature scenario for nonlinear rheology, Phys. Rev. E {\bf 61}, 5464-5472 (2000).\\
Bonn, D, Tanaka, H., Wegdam, G. H., Kellay, H. and, Meunier, J., Aging of a colloidal Wigner glass, Europhys. Lett. {\bf 45}, 52-57 (1999).\\
Bonn, D., Kellay, H., Tanaka, H., Wegdam, G. H. and, Meunier, J., Laponite: what is the difference between a gel and a glass, Langmuir {\bf 15}, 7534-7536 (1999).\\
Bonn, D., Tanase, S., Abou, B., Tanaka, H., and  Meunier, J., Laponite: Aging and shear rejuvenation of a colloidal glass, Phys. Rev. Lett. {\bf 89}, 015701 1-4 (2002).\\
Bonn, D, Coussot, P., Huynh, H. T., Bertrand, F., and Debr\'egas, G., Rheology of soft glassy materials, Europhys. Lett. {\bf 59}, 786-792 (2002).\\
Bouchaud, J.-P., Weak ergodicity breaking and aging in disordered systems, J. Phys. I (France) {\bf 2}, 1705-1713 (1992).\\
Bouchaud, J.-P., Cugliandolo, L., Kurchan, J. and  Mézard, M., Mode-coupling approximations, glass theory and disordered systems, Physica A {\bf 226}, 243-273 (1996).\\
Brady, J, and Foss, D. R., Structure, diffusion and rheology of Brownian suspensions by Stokesian Dynamics simulation, J. Fluid Mech. {\bf 407}, 167-200 (2000).\\
Coussot, P., Nguyen, Q. D., Huynh, H. T., and Bonn, D., Avalanche behavior in yield stress fluids, Phys. Rev. Lett. {\bf 88}, 175501 1-4 (2002).\\
Coussot, P., Nguyen, Q. D., Huynh, H. T., and Bonn, D., Viscosity bifuraction in thixotropic, yielding fluids, J. Rheol. {\bf 46}, 573-589 (2002).\\ 
Cugliandolo, L. F., Kurchan, J. and Peliti, L., Energy flow, partial equilibration, and effective temperatures in systems with slow dynamics, Phys. Rev. E {\bf 55}, 3898-3914 (1997).\\
Cugliandolo, L. F., Kurchan, J., Le Doussal, P., and  Peliti, L., Glassy behaviour in disordered systems with nonrelaxational dynamics, Phys. Rev. Lett. {\bf 78}, 350-353 (1997).\\
Dawson, K., Foffi, G., Fuchs, M., G\"{o}tze, W., Sciortino, F., Sperl, M., Tartaglia, P., Voigtmann, Th., and Zaccarelli, E., Higher order glass-transition singularities in colloidal systems with attractive interactions, Phys. Rev. E., {\bf 63}, 011401 1-17 (2001).\\ 
D\'erec, C., Ajdari, A., and Lequeux, F., Rheology and aging : a simple approach, Eur. Phys. J. E {\bf 4}, 355-361 (2001).\\
Fielding, S. M., Sollich, P., and Cates, M. E., Ageing and rheology in soft materials, J. Rheol. {\bf 44}, 323-369 (2000).\\
H\'ebraud, P., and Lequeux, F., Mode-coupling theory for the pasty rheology of soft glassy materials, Phys. Rev. Lett. {\bf 81}, 2934-2937 (1998).\\
Horner, H., Drift, creep and pinning of a particle in a correlated random potential, Z. Physik B {\bf 100}, 243-257 (1996). \\
Kob, W. and Barrat, J., Aging Effects in a Lennard-Jones Glass, Phys. Rev. Lett. {\bf 78}, 4581-4584 (1997).\\
Kroon, M., Vos, W. L., and  Wegdam, G. H., Structure and formation of a gel of colloidal disks, Phys. Rev. E {\bf 57}, 1962-1970 (1998).\\
Kurchan, J., Rheology, and how to stop aging, in  Edwards, S. F., Liu, A., and Nagel, R. S., (Editors), Jamming and Rheology: Constrained Dynamics on Microscopic 
and Macroscopic Scales (London, 2001); Preprint available on  http://xxx.lpthe.jussieu.fr/abs/cond-mat/9812347. \\
Larson, R. G., The structure and rheology of complex fluids (Oxford University Press, New York, 1999).\\
Levitz, P., Lecolier, E., Mourchid, A., Delville A., and, Lyonnard, S., Liquid-solid transition of Laponite suspensions at very low ionic strength: long-range electrostatic stabilisation of anisotropic colloids, Europhys. Lett. {\bf 49}, 672-677 (2000).\\
Melrose, J. R., Ball, R. C., and Silbert, L. E., The rheology and microstructure of concentrated aggregated colloids, J. Rheol. {\bf 43}, 673-700 (1999);\\ 
Melrose, J. R., Ball, R. C., and Silbert, L. E., A structural analysis of concentrated colloids under flow, Mol. Phys. {\bf 96}, 1667-1675 (1999).\\
Melrose, J. R., Ball, R. C., The pathological behaviour of sheared hard spheres with hydrodynamic interactions, Europhys. Lett. {\bf 32}, 535-540 (1995).\\
Monthus, C., and Bouchaud, J.-P., Models of trap and glass phenomenology, J. Phys. A {\bf 29}, 3847-3869 (1996).\\
Mourchid, A., and P. Levitz, P., Long-term gelation of Laponite aqueous dispersions, Phys. Rev. E {\bf 57}, 4887-4890 (1998).\\
Nelson, P., and  Allen, G. D., (Editors), Transport theory and Statistical Physics : relaxation kinetics in Supercooled liquids Mode-coupling theory and its experimental tests (Marcel Dekker, New york, 1995).\\
Nicola\"{\i}, T., and Cocard, S., Structure of gels and aggregates of disk-like colloids, Eur. Phys. J. E {\bf 5}, 221-227 (2001).\\
Pignon, F., Piau, P., and  Magnin, A., Butterfly Light Scattering Pattern and Rheology of a Sheared Thixotropic Clay Gel, Phys. Rev. Lett. {\bf 79}, 4689-4692 (1997).\\
Pusey, P. N., and  Van Megen, W., Observation of a glass transition in suspensions of spherical colloidal particles, Phys. Rev. Lett. {\bf 59}, 2083-2086 (1987).\\
Sollich, P., Lequeux, F., Hebraud, P., and Cates, M., Rheology of soft glassy materials, Phys. Rev. Lett. {\bf 78}, 2020-2023 (1997).\\
Sollich, P., Rheological constitutive equation for a model of soft glassy materials, Phys. Rev. E {\bf 58}, 738-759 (1998).\\
Thompson, D. W., and  Butterworth, J. T., The nature of Laponite and its aqueous dispersions, J. Colloid Interface Sci. {\bf 151}, 236-243 (1992).\\
Yamamoto, R., and Onuki, A., Dynamics of highly supercooled liquids: Heterogeneity, rheology, and diffusion, Phys. Rev E. {\bf 58}, 3515-3529 (1998).

\end{document}